\documentclass[eqsecnum,twocolumn,aps,epsf,showpacs]{revtex4}
\usepackage{graphicx}
\usepackage{color}
\begin{document}
\draft

\title{Muon-Spin-Relaxation Study of the Cu-Spin Dynamics in Electron-Doped High-$T_{\mathrm{c}}$ Superconductor Pr$_{0.86}$LaCe$_{0.14}$Cu$_{1-y}$Zn$_y$O$_4$}

\author{Risdiana}
\altaffiliation[Present address: ]{Advanced Meson Science Laboratory, Nishina Center for Accelerator-Based Science, The Institute of Physical and Chemical Research (RIKEN), 2-1 Hirosawa, Wako 351-0198, Japan}
\altaffiliation[Permanent address: ]{Department of Physics, Faculty of Mathematics and Natural Sciences, Padjadjaran University, Jl. Raya Bandung-Sumedang Km. 21, Jatinangor 45363, Indonesia}
\author{T. Adachi}
\thanks{Corresponding author: adachi@teion.apph.tohoku.ac.jp}
\author{N. Oki}
\author{Y. Koike}
\affiliation{Department of Applied Physics, Graduate School of Engineering, Tohoku University, 6-6-05 Aoba, Aramaki, Aoba-ku, Sendai 980-8579, Japan}

\author{T. Suzuki}
\altaffiliation[Present address: ]{College of Liberal Arts, International Christian University, 3-10-2 Osawa, Mitaka, Tokyo 181-8585, Japan}
\author{I. Watanabe}
\affiliation{Advanced Meson Science Laboratory, Nishina Center for Accelerator-Based Science, The Institute of Physical and Chemical Research (RIKEN), 2-1 Hirosawa, Wako 351-0198, Japan}

\date{\today}

\begin{abstract}
Muon-spin-relaxation ($\mu$SR) measurements have been performed for the partially Zn-substituted electron-doped high-$T_{\mathrm{c}}$ superconductor Pr$_{0.86}$LaCe$_{0.14}$Cu$_{1-y}$Zn$_y$O$_{4+\alpha-\delta }$ with $y$ = 0 - 0.05 and the reduced oxygen content $\delta$ = 0 - 0.09, in order to investigate nonmagnetic Zn-impurity effects on the Cu-spin dynamics. For all the measured samples with 0.01 $\leq$ $\delta$ $\leq$ 0.09, it has been found that a fast depolarization of muon spins is observed below 100 K due to the effect of Pr$^{3+}$ moments and that the $\mu$SR time spectrum in the long-time region above 5 $\mu$sec increases with decreasing temperature at low temperatures below 30 K possibly due to slowing down of the Cu-spin fluctuations assisted by Pr$^{3+}$ moments. No Zn-induced slowing down of the Cu-spin fluctuations has been observed for moderately oxygen-reduced samples with 0.04 $\le$ $\delta$ $\le$ 0.09, which is very different from the $\mu$SR results of La$_{2-x}$Sr$_{x}$Cu$_{1-y}$Zn$_{y}$O$_4$. The possible reason may be that there are no dynamical stripe correlations of spins and electrons in the electron-doped high-$T_{\mathrm{c}}$ cuprates or that the effect of Pr$^{3+}$ moments on the $\mu$SR spectra is stronger than that of a small amount of Zn impurities.   
\end{abstract}
\vspace*{2em}
\pacs{74.62.Dh, 74.72.Ek, 76.75.+i.}
\maketitle
\newpage

\section{Introduction}\label{intro}
The so-called hole-electron doping symmetry in the high-$T_{\mathrm{c}}$ cuprates has been one of central interests in relation to the mechanism of the high-$T_{\mathrm{c}}$ superconductivity. 
Phase diagrams of the hole- and electron-doped systems are very similar to each other. That is, the parent compounds are both Mott insulators exhibiting long-range antiferromagnetic order with similar values of the N\'{e}el temperature. The superconducting phases appear through doping holes or electrons into the Mott insulators. These properties lead to the view of hole-electron doping symmetry. 

On the other hand, some properties in the electron-doped superconductors have been found to be different from those in the hole-doped superconductors, leading to the hole-electron doping asymmetry. First, the effectiveness of carriers for destroying the long-range antiferromagnetic order is different between the hole- and electron-doped systems. In the electron-doped system of Nd$_{2-x}$Ce$_x$CuO$_4$, the long-range antiferromagnetic order survives up to $x$ $\sim$ 0.13 \cite{takagi}, while it survives only up to $x$ $\sim$ 0.02 in the hole-doped system of La$_{2-x}$Sr$_x$CuO$_4$. 
Second, in the inelastic neutron-scattering measurements, an incommensurate spin-correlation, which may be due to the so-called dynamically fluctuating stripes of spins and holes \cite{tranquada}, has been found in the hole-doped system \cite{yamada1}. In the electron-doped system, on the other hand, a commensurate spin-correlation, which is related to the simple antiferromagnetic order, has been observed \cite{yamada}.
As for impurity effects, different behaviors between the hole- and electron-doped systems are also observed. For examples, the superconductivity in the electron-doped system is suppressed through the substitution of magnetic Ni for Cu more markedly than through the substitution of nonmagnetic Zn for Cu \cite{tarascon}, which is contrary to the result in the hole-doped system \cite{xiao}.

From the view point of effects of nonmagnetic impurities on the Cu-spin dynamics, formerly, we have performed zero-field (ZF) muon-spin-relaxation ($\mu$SR) measurements in a wide range of hole concentration of $x$ = 0.10 - 0.30 in the hole-doped high-$T_{\mathrm{c}}$ superconductor La$_{2-x}$Sr$_x$Cu$_{1-y}$Zn$_y$O$_4$ \cite{adachi3,nabe1,risdi,adachi4}. It has been found that a slight amount of Zn tends to induce slowing down of the Cu-spin fluctuations in the whole superconducting regime, which is able to be interpreted as being due to pinning and stabilization of the dynamically fluctuating stripes of spins and holes. This result might point to the importance of the dynamical stripe correlations in the appearance of high-$T_{\mathrm{c}}$ superconductivity in the hole-doped system \cite{kivelson}. 

In this paper, we investigate Zn-impurity effects on the Cu-spin dynamics  in the electron-doped high-$T_{\mathrm{c}}$ superconductor Pr$_{0.86}$LaCe$_{0.14}$Cu$_{1-y}$Zn$_y$O$_{4+\alpha-\delta }$ with $y$ = 0 - 0.05 and the reduced oxygen content $\delta$ = 0 - 0.09 from the ZF- and longitudinal-field (LF) $\mu$SR measurements, in order to elucidate whether or not the concept of the pinning of the dynamical stripe correlations by Zn holds good in the electron-doped system \cite{risdi2,risdi3}.

\section{Experimental}
Polycrystalline samples of the electron-doped Pr$_{0.86}$LaCe$_{0.14}$Cu$_{1-y}$Zn$_y$O$_{4+{\alpha-\delta}}$ with $y$ = 0, 0.01, 0.02, 0.05 were prepared by the ordinary solid-state reaction method as follows \cite{koikejjap}. Raw materials of dried La$_2$O$_3$, Pr$_6$O$_{11}$, CeO$_2$, CuO and ZnO powders were mixed in a stoichiometric ratio and prefired in air at 900$^{\rm o}$C for 20 h. The prefired materials were reground and pressed into pellets of 10 mm in diameter, and sintered in air at 1100$^{\rm o}$C for 16 h with repeated regrinding. As-grown samples of Pr$_{0.86}$LaCe$_{0.14}$Cu$_{1-y}$Zn$_y$O$_{4+{\alpha}}$ ($|\alpha| \ll 1$) were post-annealed in flowing Ar gas of high purity (6N) at various temperatures in a range of 900$^{\rm o}$C - 960$^{\rm o}$C for 8 h - 12 h in order to remove the excess oxygen at the so-called apical site. The value of $\delta$ was estimated from the weight change before and after annealing.
All of the samples were checked by the powder x-ray diffraction measurements to be of the single phase. 
Electrical-resistivity and DC magnetic-susceptibility measurements were also carried out to check the superconducting transition temperature, $T_{\mathrm{c}}$, and the quality of the samples, which was found to be good. The ZF- and LF-$\mu$SR measurements were performed at low temperatures down to 0.3 K at the RIKEN-RAL Muon Facility at the Rutherford-Appleton Laboratory in the UK using a pulsed positive surface muon beam.

\section{Results}
The superconductivity in the electron-doped system of Pr$_{0.86}$LaCe$_{0.14}$Cu$_{1-y}$Zn$_y$O$_{4+{\alpha-\delta}}$ is affected by the value of $\delta$, so that the samples are grouped into four classes with different $\delta$ values: as-grown ($\delta$ = 0), very small $\delta$ ($\delta$ $<$ 0.01), small $\delta$ (0.01 $\leq$ $\delta$ $<$ 0.04) and large $\delta$ values (0.04 $\leq$ $\delta$ $\leq$ 0.09).   

\begin{figure}[tbp]
\begin{center}
\includegraphics[width=0.8\linewidth]{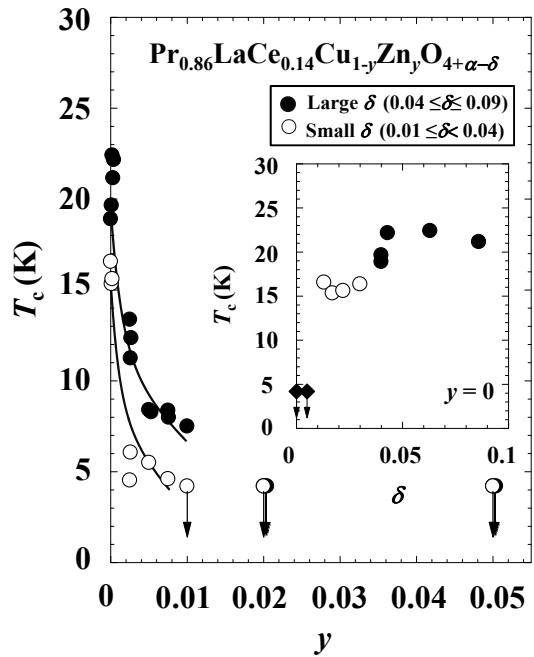}
\end{center}
\caption{Dependence of $T_{\mathrm{c}}$ on the Zn-concentration $y$ of Pr$_{0.86}$LaCe$_{0.14}$Cu$_{1-y}$Zn$_y$O$_{4+{\alpha-\delta}}$ with small $\delta$ (0.01 $\leq$ $\delta$ $<$ 0.04) and large $\delta$ values (0.04 $\leq$ $\delta$ $\leq$ 0.09). The inset shows the dependence of $T_{\mathrm{c}}$ on $\delta$ for Zn-free samples of $y$ = 0. Arrows indicate non-superconducting samples above 4.2 K.}  
\label{fig:fig1} 
\end{figure}

Figure 1 shows the dependence of $T_{\mathrm{c}}$ on the Zn-concentration $y$ in Pr$_{0.86}$LaCe$_{0.14}$Cu$_{1-y}$Zn$_y$O$_{4+{\alpha-\delta}}$ with small $\delta$ and large $\delta$ values. The $T_{\mathrm{c}}$ is defined as the temperature where the resistivity drops to 50 $\%$ of the normal-state value. It is found that $T_{\mathrm{c}}$ decreases with increasing $y$, indicating that Zn is well substituted for Cu. The samples with large $\delta$ values show superconductivity below $y$ = 0.02, while they are non-superconducting above $y$ = 0.02 at temperatures above 4.2 K. On the other hand, the samples with small $\delta$ values are non-superconducting above $y$= 0.01 at temperatures above 4.2 K. As for the Zn-free Pr$_{0.86}$LaCe$_{0.14}$CuO$_{4+{\alpha-\delta}}$, $T_{\mathrm{c}}$ increases with increasing $\delta$, as shown in the inset of Fig. 1. The samples with small and large $\delta$ values show superconductivity with $T_{\mathrm{c}}$ ranging from 15 K to 17 K (the average $T_{\mathrm{c}}$ $\sim$ 16 K) and from 18 K to 22 K (the average $T_{\mathrm{c}}$ $\sim$ 20 K), respectively, while the samples with $\delta$ = 0 and very small $\delta$ values are non-superconducting above 4.2 K. 

\begin{figure}[tbp]
\begin{center}
\includegraphics[width=0.8\linewidth]{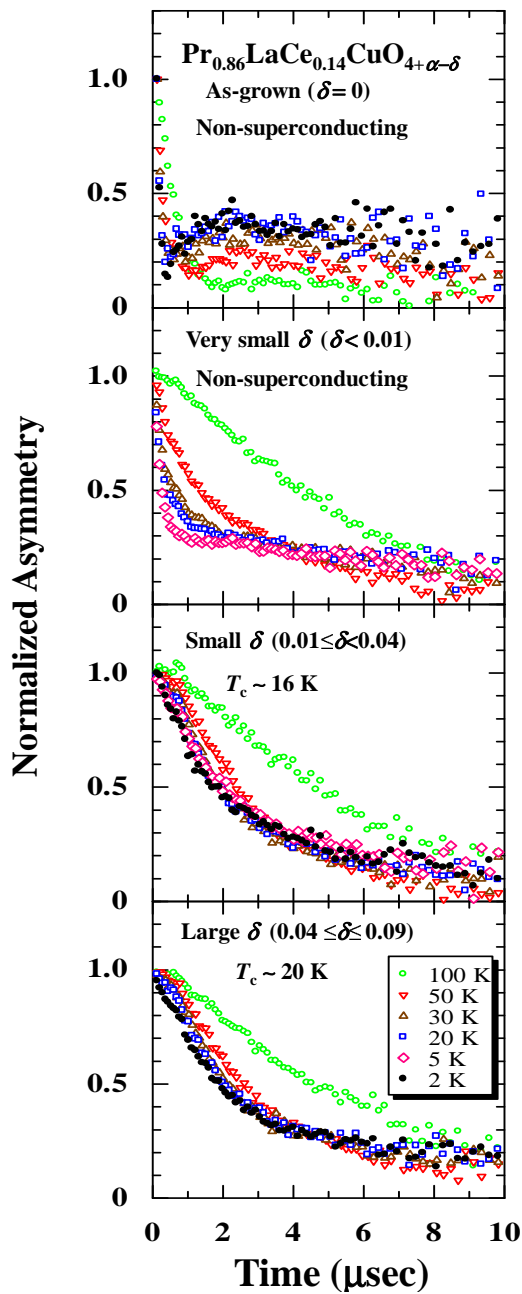}
\end{center}
\caption{(color online) ZF-$\mu$SR time spectra of the Zn-free Pr$_{0.86}$LaCe$_{0.14}$CuO$_{4+{\alpha-\delta}}$ with different $\delta$ values at various temperatures.}  
\label{fig:fig2} 
\end{figure}

Figure 2 shows the ZF-$\mu$SR time spectra of the Zn-free Pr$_{0.86}$LaCe$_{0.14}$CuO$_{4+{\alpha-\delta}}$ with $\delta$ = 0, very small $\delta$, small $\delta$ and large $\delta$ values. 
For the as-grown sample with $\delta$ = 0, a muon-spin precession is observed even at a high temperature of 100 K due to the formation of a long-range antiferromagnetic order. For the samples with very small $\delta$, small $\delta$ and large $\delta$ values, a Gaussian-like depolarization is observed at high temperatures above $\sim$ 100 K due to randomly oriented nuclear spins, and an exponential-like depolarization of muon spins is observed at low temperatures below $\sim$ 50 K. For the sample with the very small $\delta$ value, a muon-spin precession due to the formation of a long-range antiferromagnetic order is observed at low temperatures below $\sim$ 5 K. For the samples with small and large $\delta$ values, on the other hand, the time spectra in the short-time region below 5 $\mu$sec are almost independent of temperature at low temperatures below 30 K and no muon-spin precession is observed, indicating the absence of any long-range magnetic order above 2 K. 
As for the spectra at high temperatures above 30 K, the temperature-dependent change of the spectra is independent of $\delta$ for superconducting samples with small and large $\delta$ values. Taking into account the fact that the Cu-spin dynamics strongly depends on $\delta$ as well as the Ce concentration \cite{kuroshima}, the temperature-dependent change of the spectra above 30 K in both small and large $\delta$ values is not regarded as being due to the Cu-spin dynamics but due to static random magnetism of small magnetic moments of Pr$^{3+}$ ions induced by the mixing of the excited state in the crystal electric field \cite{kadono, kadono2}. 
\begin{figure}[tbp]
\begin{center}
\includegraphics[width=0.9\linewidth]{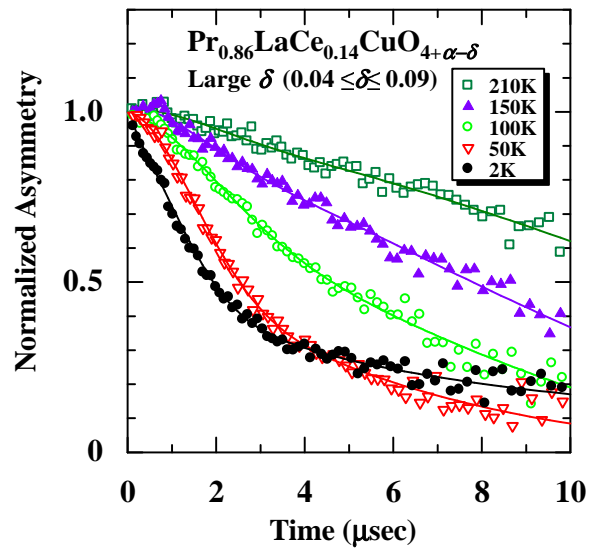}
\end{center}
\caption{(color online) ZF-$\mu$SR time spectra of the Zn-free Pr$_{0.86}$LaCe$_{0.14}$CuO$_{4+{\alpha-\delta}}$ with large $\delta$ values at various temperatures. Solid lines are the best-fit results using the two-component function, $A(t)$  = $A_{\mathrm{s}}$$\mathrm{exp}(-(\lambda t)^ \beta )$ + $A_{\mathrm{G}}$$\mathrm{exp}(-\sigma ^2 t^2)$.}  
\label{fig:fig3} 
\end{figure}
Watching the spectra of the samples with large $\delta$ values in detail, it is found that the normalized asymmetry in the long-time region above 5 $\mu$sec increases with decreasing temperature at low temperatures, as clearly shown in Fig. 3. The behavior is analogous to the recovery of the normalized asymmetry to 1/3 in the long-time region usually observed in a long-range magnetically ordered state. The fast depolarization with no muon-spin precession is considered to be due to the formation of not a long-range but a short-range magnetic order. Therefore, it is suggested that the Cu-spin correlation is developed and the Cu-spin fluctuations exhibit slowing down at low temperatures for the samples with large $\delta$ values.

\begin{figure}[tbp]
\begin{center}
\includegraphics[width=0.9\linewidth]{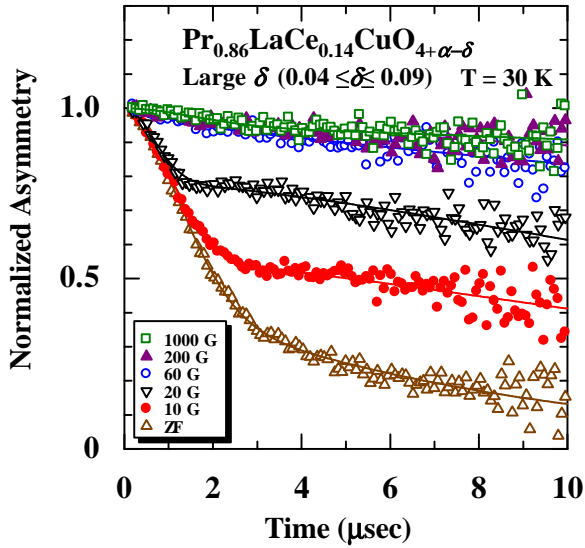}
\end{center}
\caption{(color online) LF-$\mu$SR time spectra in the normal state of the Zn-free Pr$_{0.86}$LaCe$_{0.14}$CuO$_{4+{\alpha-\delta}}$ with large $\delta$ values at 30 K under various LF up to 1000 G. Solid lines are the best-fit results using the two-component function, $A(t) = A_{\mathrm{s}} {\mathrm{exp}}(-(\lambda t)^ \beta) + A_{\mathrm{G}} {\mathrm{exp}}(-\sigma ^2 t^2)$.}  
\label{fig:fig4} 
\end{figure}

In order to clarify the dual existence of the static magnetism and dynamical spin-fluctuations, LF-$\mu$SR measurements were performed under LF up to 1000 G. Figure 4 shows the LF-$\mu$SR time spectra in the normal state of the Zn-free Pr$_{0.86}$LaCe$_{0.14}$CuO$_{4+{\alpha-\delta}}$ with large $\delta$ values at 30 K. The tail of the spectrum is found to be gradually quenched with increasing LF up to 60 G, suggesting the existence of static magnetic moments. However a long-time depolarization is still observed even in high LF of 1000 G, suggesting that dynamically fluctuating internal fields still exist at the muon site. Therefore, the LF-$\mu$SR results suggest the coexistence of the static magnetism and dynamical spin fluctuations in the sample of Pr$_{0.86}$LaCe$_{0.14}$CuO$_{4+{\alpha-\delta}}$ with large $\delta$ values. 

\begin{figure}[tbp]
\begin{center}
\includegraphics[width=0.8\linewidth]{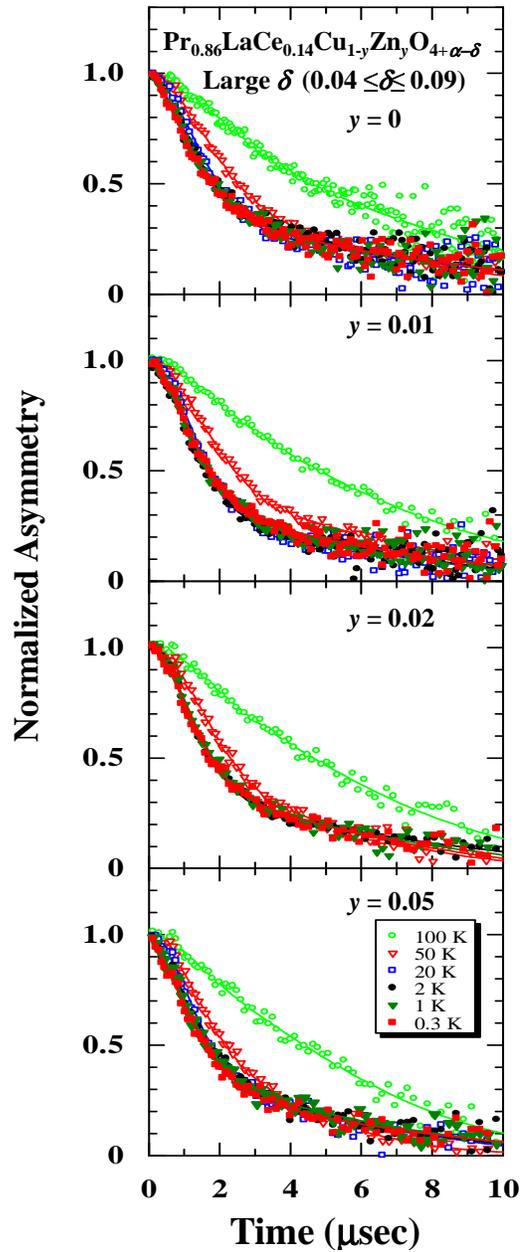}
\end{center}
\caption{(color online) ZF-$\mu$SR time spectra of Pr$_{0.86}$LaCe$_{0.14}$Cu$_{1-y}$Zn$_y$O$_{4+{\alpha-\delta}}$ with $y$ = 0 - 0.05 and large $\delta$ values at low temperatures down to 0.3 K. Solid lines are the best-fit results using the two-component function, $A(t) = A_{\mathrm{s}} {\mathrm{exp}}(-(\lambda t)^ \beta) + A_{\mathrm{G}} {\mathrm{exp}}(-\sigma ^2 t^2)$.}  
\label{fig:fig5} 
\end{figure}

Figure 5 shows the ZF-$\mu$SR time spectra of the Zn-substituted Pr$_{0.86}$LaCe$_{0.14}$Cu$_{1-y}$Zn$_y$O$_{4+{\alpha-\delta}}$ with $y$ = 0 - 0.05 and large $\delta$ values. It is found that the spectra are independent of the Zn concentration. That is, Zn-induced slowing down of the Cu-spin fluctuations as observed in the hole-doped system is not observed.

In the analysis of the ZF-$\mu$SR time spectra, we assume that the implanted muons occupy the site near the edge of tetrahedron configuration with Pr/La atoms at their corner \cite{kadono}. Since the muon feels an internal field from two kinds of moments addressed to Pr$^{3+}$ and Cu$^{2+}$ moments, the effect of both moments should be taken into account. Among several equations such as dynamic Kubo-Toyabe function, it has been found that the ZF-$\mu$SR time spectra are best fitted with the following two-component function: 
$$
\; \; \; \; A(t) = A_{\mathrm{s}} {\mathrm{exp}}(-(\lambda t)^ \beta) + A_{\mathrm{G}} {\mathrm{exp}}(-\sigma ^2 t^2) \; \; \; \; \; \; \; \; \; \; \; \; \; \; \; \; (1)         
$$
The first term represents a stretched-exponential component in a region where effects of nuclear spins and Cu spins are dominant. $A_{\mathrm{s}}$, $\lambda$ and $\beta$ are the initial asymmetry, the relaxation rate of muon spins and the power of damping, respectively. The second term represents a static Gaussian component in a region where the small Pr$^{3+}$ moments are dominant. $A_{\mathrm{G}}$ is the initial asymmetry and $\sigma$ is due to the distribution width of dipolar fields at the muon site. The increase in $A(t)$ observed in the long-time region at low temperatures is reflected by the increase in $A_{\mathrm{s}}$. The time spectra are well fitted with this function, as clearly shown in Figs. $3-5$.

\begin{figure}[tbp]
\begin{center}
\includegraphics[width=0.7\linewidth]{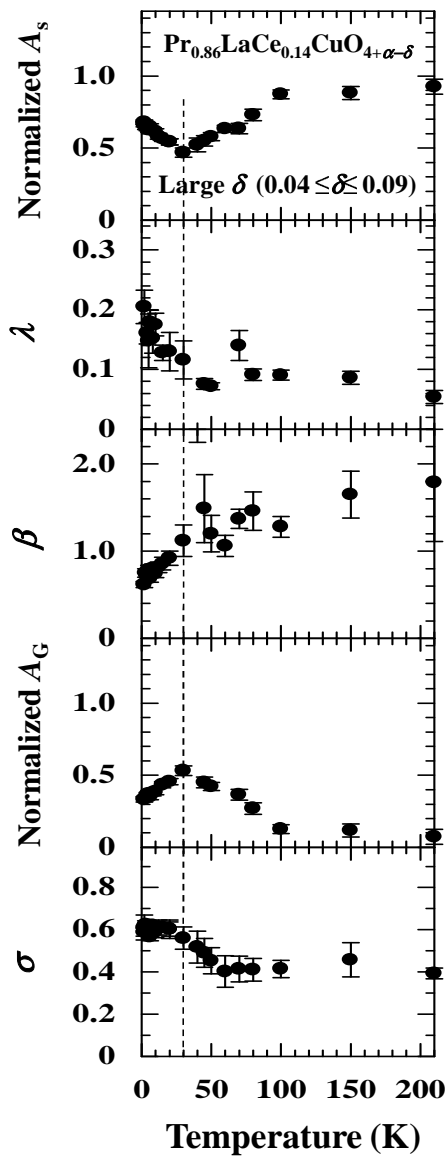}
\end{center}
\caption{Temperature dependence of fitting parameters in the function, $A(t)$ = $A_{\mathrm{s}}$$\mathrm{exp}(-(\lambda t)^ \beta )$ + $A_{\mathrm{G}}$$\mathrm{exp}(-\sigma ^2 t^2)$, for Pr$_{0.86}$LaCe$_{0.14}$CuO$_{4+{\alpha-\delta}}$ with large $\delta$ values. $A_{\mathrm{s}}$ : initial asymmetry of the stretched-exponential component. $\lambda$ :  relaxation rate of muon spins. $\beta$ : power of the damping of the stretched-exponential component. $A_{\mathrm{G}}$ : initial asymmetry  of the static Gaussian component. $\sigma$ : distribution width of dipolar fields at the muon site.}  
\label{fig:fig6} 
\end{figure}

Figure 6 shows the temperature dependence of fitting parameters $A_{\mathrm{s}}$, $\lambda$, $\beta$, $A_{\mathrm{G}}$ and $\sigma$ for the Zn-free Pr$_{0.86}$LaCe$_{0.14}$CuO$_{4+{\alpha-\delta}}$ with large $\delta$ values. At high temperatures above 100 K, all parameters are almost independent of temperature, indicating no other effect than that of nuclear spins. Below 100 K, $A_{\mathrm{G}}$ starts to increase with decreasing temperature down to 30 K, accompanied by an increase in $\sigma$ below $\sim$ 60 K. This is regarded as being due to development of the static random magnetism of the small Pr$^{3+}$ moments. Below 30 K, $A_{\mathrm{G}}$ decreases and $A_{\mathrm{s}}$ and $\lambda$ increase with decreasing temperature. This is regarded as being due to the development of the Cu-spin correlation. In fact, $\beta$ tends to fall down to 0.5 at 0.3 K, suggesting the formation of a spin-glass state of Cu spins \cite{cywinski}. These results indicate that, with decreasing temperature, the development of the static random magnetism of the small Pr$^{3+}$ moments starts to appear below 100 K and the development of the Cu-spin correlation, namely, the slowing down of the Cu-spin fluctuations starts to appear below 30 K.

\begin{figure}[tbp]
\begin{center}
\includegraphics[width=1.0\linewidth]{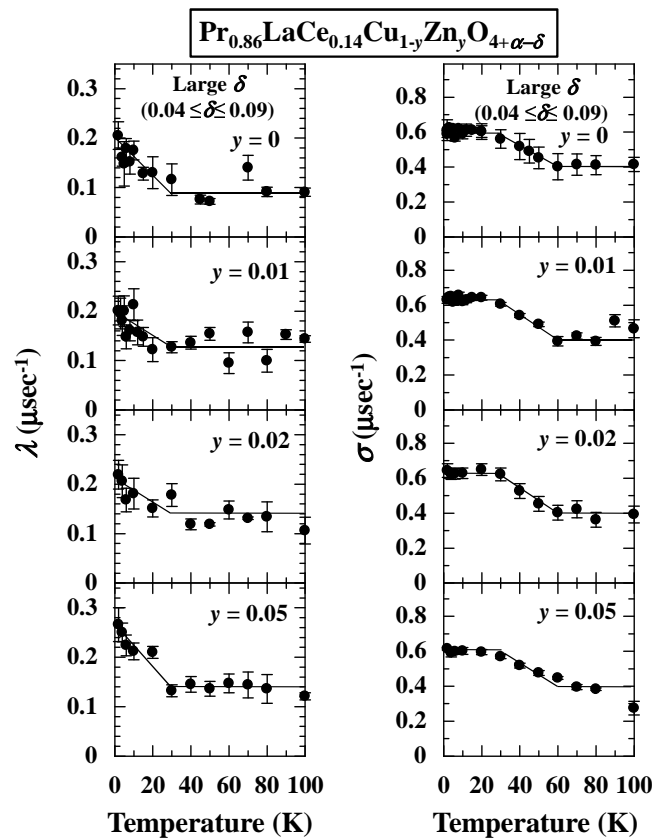}
\end{center}
\caption{Temperature dependence of $\lambda$ and $\sigma$ for Pr$_{0.86}$LaCe$_{0.14}$Cu$_{1-y}$Zn$_y$O$_{4+{\alpha-\delta}}$ with large $\delta$ values. Solid lines are to guide the reader's eye.}  
\label{fig:fig7} 
\end{figure}

Figure 7 shows the temperature dependence of $\lambda$ and $\sigma$ of the Zn-substituted Pr$_{0.86}$LaCe$_{0.14}$Cu$_{1-y}$Zn$_y$O$_{4+{\alpha-\delta}}$ with large $\delta$ values. For all the Zn-substituted samples, $\lambda$ increases with decreasing temperature below $\sim$ 30 K due to the slowing down of the Cu-spin fluctuations. On the other hand, $\sigma$ increases below $\sim$ 60 K and is saturated below $\sim$ 30 K due to the development of the static random magnetism of the small Pr$^{3+}$ moments. For both $\lambda$ and $\sigma$, it is found that no significantly different behavior is observed between the samples with different $y$ values, indicating no Zn-induced development of the Cu-spin correlation, namely, no Zn-induced slowing down of the Cu-spin fluctuations. This result is very different from that observed in the hole-doped La$_{2-x}$Sr$_{x}$Cu$_{1-y}$Zn$_y$O$_4$ \cite{adachi3,nabe1,risdi,adachi4}.

\section{Discussion}
One important feature in the present results is that the trace of the development of the Cu-spin correlation is observed at low temperatures below $\sim$ 30 K even for the Zn-free sample. Fujita $et$ $al.$ have also reported the possible development of the Cu-spin correlation and the slowing down of the Cu-spin fluctuations in the single-crystal Pr$_{1-x}$LaCe$_{x}$CuO$_{4+{\alpha-\delta}}$ with $x$ = 0.11 undergoing the  appearance of static/quasi static internal magnetic fields at the muon site \cite{fujita2}. The possible origin of the slowing down observed even in the Zn-free samples is enhancement of the Cu-spin correlation assisted by Pr$^{3+}$ moments. 

The other important feature is no Zn-induced slowing down of the Cu-spin fluctuations. That is, Zn impurities do not appear to affect the Cu-spin dynamics, which is very different from the results of the hole-doped high-$T_{\mathrm{c}}$ cuprates \cite{adachi3,nabe1,risdi,adachi4}. This may be understood in two ways. First, there may be no dynamically fluctuating stripes of spins and electrons in the electron-doped system \cite{yamada,fujita}, because the existence of the dynamical stripes leads to the Zn-induced slowing down of the Cu-spin fluctuations in the hole-doped system. Secondly, the effect of Pr$^{3+}$ moments is stronger than that of a small amount of Zn impurities. In any case, the Cu-spin dynamics observed within the $\mu$SR time window (10$^{-6}$ - 10$^{-11}$ sec) in the electron-doped cuprates may not be so affected by any impurity as follows. Generally, impurities tend to make carriers localized. Since doped holes give rise to magnetic frustration between Cu spins in the hole-doped cuprates, whether holes are mobile or localized may affect the Cu-spin dynamics strongly, leading to large effects of impurities on the Cu-spin dynamics. In the electron-doped cuprates, on the other hand, doped electrons give rise to no magnetic frustration between Cu spins and give rise to only dilution of Cu spins. Therefore, whether doped electrons are mobile or localized may not affect the Cu-spin dynamics so much, leading to no significant effects of impurities on the Cu-spin dynamics.

\section{Summary}
We have investigated the Cu-spin dynamics from ZF-$\mu$SR measurements in the partially Zn-substituted electron-doped high-$T_{\mathrm{c}}$ superconductor Pr$_{0.86}$LaCe$_{0.14}$Cu$_{1-y}$Zn$_y$O$_{4+{\alpha-\delta}}$ changing the reduced oxygen content $\delta$ ($\delta$ $\le$ 0.09) and $y$  ($y$ $\le$ 0.05). 
For the Zn-free samples of Pr$_{0.86}$LaCe$_{0.14}$CuO$_{4+{\alpha-\delta}}$ with small $\delta$ (0.01 $\leq$ $\delta$ $<$ 0.04) and large $\delta$ values (0.04 $\leq$ $\delta$ $\leq$ 0.09), it has been found that a fast depolarization of muon spins is observed below 100 K due to the effect of Pr$^{3+}$ moments and that the $\mu$SR time spectrum in the long-time region above 5 $\mu$sec increases with decreasing temperature at low temperatures below 30 K. The latter suggests the possible slowing down of the Cu-spin fluctuations assisted by Pr$^{3+}$ moments. The dual existence of the static magnetism of Pr$^{3+}$ moments and the Cu-spin fluctuations has been confirmed from LF-$\mu$SR measurements. Moreover, no Zn-induced slowing down of the Cu-spin fluctuations has been observed for moderately oxygen-reduced samples with 0.04 $\le$ $\delta$ $\le$ 0.09, which is very different from the $\mu$SR results of the hole-doped high-$T_{\mathrm{c}}$ superconductor La$_{2-x}$Sr$_{x}$Cu$_{1-y}$Zn$_{y}$O$_4$. Possible reasons are as follows: (i) There may be no dynamical stripe correlations of spins and electrons in the electron-doped system. (ii) The effect of Pr$^{3+}$ moments on the $\mu$SR spectra may be stronger than that of a small amount of Zn impurities.

\section*{Acknowledgments}
We would like to thank K. Hachitani for his technical support in a part of the $\mu$SR measurements. 
This work was partly supported by Joint Programs of the Japan Society for the Promotion of Science and by a Grant-in-Aid for Scientific Research from the Ministry of Education, Culture, Sport, Science and Technology, Japan.

\end{document}